\documentclass[final,5p,authoryear,times]{elsarticle}
\pdfoutput=1

\usepackage{graphicx}
\usepackage{amsfonts,amsmath,amssymb}
\usepackage[utf8]{inputenc}
\usepackage{textcomp}
\usepackage{enumitem} 
\usepackage{dblfloatfix} 
\usepackage[colorlinks,linkcolor=blue,citecolor=blue,urlcolor=blue,pdftitle={Crowdsourcing quality control for Dark Energy Survey images},pdfauthor={Melchior et al.}]{hyperref}
\usepackage{caption}
\usepackage{flushend}

\let\origfootnote\footnote
\renewcommand{\footnote}[1]{\kern.06em\origfootnote{#1}}
\newcommand{\punctfootnote}[1]{\kern-.06em\origfootnote{#1}}

\usepackage[shortcuts]{extdash}

\usepackage{eso-pic}

\AddToShipoutPictureBG*{%
  \AtPageUpperLeft{%
    \hspace{0.75\paperwidth}%
    \raisebox{-3.5\baselineskip}{%
      \makebox[0pt][l]{\textnormal{DES 2015-0109}}
}}}%

\AddToShipoutPictureBG*{%
  \AtPageUpperLeft{%
    \hspace{0.75\paperwidth}%
    \raisebox{-4.5\baselineskip}{%
      \makebox[0pt][l]{\textnormal{FERMILAB-PUB-15-487}}
}}}%

\DeclareRobustCommand{\object}[1]{%
   #1%
}

\journal{Astronomy \& Computing}
\begin{document}
\begin{frontmatter}

\title{Crowdsourcing quality control for Dark Energy Survey images}

\author[1,2]{P.~Melchior}
\ead{peter@pmelchior.net}
\cortext[cor1]{Corresponding author}
\author[3]{E.~Sheldon}
\author[4]{A.~Drlica-Wagner}
\author[5,6]{E.~S.~Rykoff}
\author[7]{T. M. C.~Abbott}
\author[8,9]{F.~B.~Abdalla}
\author[4]{S.~Allam}
\author[10,8,11]{A.~Benoit-L{\'e}vy}
\author[8]{D.~Brooks}
\author[4]{E.~Buckley-Geer}
\author[12,13]{A.~Carnero~Rosell}
\author[14,15]{M.~Carrasco~Kind}
\author[16,17]{J.~Carretero}
\author[16]{M.~Crocce}
\author[18,19]{C.~B.~D'Andrea}
\author[12,13]{L.~N.~da Costa}
\author[20,21]{S.~Desai}
\author[8]{P.~Doel}
\author[22,23]{A.~E.~Evrard}
\author[4]{D.~A.~Finley}
\author[4]{B.~Flaugher}
\author[4,24]{J.~Frieman}
\author[16]{E.~Gaztanaga}
\author[23]{D.~W.~Gerdes}
\author[25,26]{D.~Gruen}
\author[14,15]{R.~A.~Gruendl}
\author[1,2]{K.~Honscheid}
\author[7]{D.~J.~James}
\author[27]{M.~Jarvis}
\author[28]{K.~Kuehn}
\author[29]{T.~S.~Li}
\author[12,13]{M.~A.~G.~Maia}
\author[27]{M.~March}
\author[29]{J.~L.~Marshall}
\author[4]{B.~Nord}
\author[12,13]{R.~Ogando}
\author[30]{A.~A.~Plazas}
\author[31]{A.~K.~Romer}
\author[32]{E.~Sanchez}
\author[4]{V.~Scarpine}
\author[32,14]{I.~Sevilla-Noarbe}
\author[7]{R.~C.~Smith}
\author[4]{M.~Soares-Santos}
\author[27]{E.~Suchyta}
\author[15]{M.~E.~C.~Swanson}
\author[23]{G.~Tarle}
\author[33]{V.~Vikram}
\author[7]{A.~R.~Walker}
\author[4]{W.~Wester}
\author[23]{Y.~Zhang}

\address[1]{Center for Cosmology and Astro-Particle Physics, The Ohio State University, Columbus, OH 43210, USA}
\address[2]{Department of Physics, The Ohio State University, Columbus, OH 43210, USA}
\address[3]{Brookhaven National Laboratory, Bldg 510, Upton, NY 11973, USA}
\address[4]{Fermi National Accelerator Laboratory, P. O. Box 500, Batavia, IL 60510, USA}
\address[5]{Kavli Institute for Particle Astrophysics \& Cosmology, P. O. Box 2450, Stanford University, Stanford, CA 94305, USA}
\address[6]{SLAC National Accelerator Laboratory, Menlo Park, CA 94025, USA}
\address[7]{Cerro Tololo Inter-American Observatory, National Optical Astronomy Observatory, Casilla 603, La Serena, Chile}
\address[8]{Department of Physics \& Astronomy, University College London, Gower Street, London, WC1E 6BT, UK}
\address[9]{Department of Physics and Electronics, Rhodes University, PO Box 94, Grahamstown, 6140, South Africa}
\address[10]{CNRS, UMR 7095, Institut d'Astrophysique de Paris, F-75014, Paris, France}
\address[11]{Sorbonne Universit\'es, UPMC Univ Paris 06, UMR 7095, Institut d'Astrophysique de Paris, F-75014, Paris, France}
\address[12]{Laborat\'orio Interinstitucional de e-Astronomia - LIneA, Rua Gal. Jos\'e Cristino 77, Rio de Janeiro, RJ - 20921-400, Brazil}
\address[13]{Observat\'orio Nacional, Rua Gal. Jos\'e Cristino 77, Rio de Janeiro, RJ - 20921-400, Brazil}
\address[14]{Department of Astronomy, University of Illinois, 1002 W. Green Street, Urbana, IL 61801, USA}
\address[15]{National Center for Supercomputing Applications, 1205 West Clark St., Urbana, IL 61801, USA}
\address[16]{Institut de Ci\`encies de l'Espai, IEEC-CSIC, Campus UAB, Carrer de Can Magrans, s/n,  08193 Bellaterra, Barcelona, Spain}
\address[17]{Institut de F\'{\i}sica d'Altes Energies (IFAE), The Barcelona Institute of Science and Technology, Campus UAB, 08193 Bellaterra (Barcelona) Spain}
\address[18]{Institute of Cosmology \& Gravitation, University of Portsmouth, Portsmouth, PO1 3FX, UK}
\address[19]{School of Physics and Astronomy, University of Southampton,  Southampton, SO17 1BJ, UK}
\address[20]{Excellence Cluster Universe, Boltzmannstr.\ 2, 85748 Garching, Germany}
\address[21]{Faculty of Physics, Ludwig-Maximilians University, Scheinerstr. 1, 81679 Munich, Germany}
\address[22]{Department of Astronomy, University of Michigan, Ann Arbor, MI 48109, USA}
\address[23]{Department of Physics, University of Michigan, Ann Arbor, MI 48109, USA}
\address[24]{Kavli Institute for Cosmological Physics, University of Chicago, Chicago, IL 60637, USA}
\address[25]{Max Planck Institute for Extraterrestrial Physics, Giessenbachstrasse, 85748 Garching, Germany}
\address[26]{Universit\"ats-Sternwarte, Fakult\"at f\"ur Physik, Ludwig-Maximilians Universit\"at M\"unchen, Scheinerstr. 1, 81679 M\"unchen, Germany}
\address[27]{Department of Physics and Astronomy, University of Pennsylvania, Philadelphia, PA 19104, USA}
\address[28]{Australian Astronomical Observatory, North Ryde, NSW 2113, Australia}
\address[29]{George P. and Cynthia Woods Mitchell Institute for Fundamental Physics and Astronomy, and Department of Physics and Astronomy, Texas A\&M University, College Station, TX 77843,  USA}
\address[30]{Jet Propulsion Laboratory, California Institute of Technology, 4800 Oak Grove Dr., Pasadena, CA 91109, USA}
\address[31]{Department of Physics and Astronomy, Pevensey Building, University of Sussex, Brighton, BN1 9QH, UK}
\address[32]{Centro de Investigaciones Energ\'eticas, Medioambientales y Tecnol\'ogicas (CIEMAT), Madrid, Spain}
\address[33]{Argonne National Laboratory, 9700 South Cass Avenue, Lemont, IL 60439, USA}

\begin{abstract}
We have developed a crowdsourcing web application for image quality control employed by the Dark Energy Survey. 
Dubbed the ``DES exposure checker'', it renders science-grade images directly to a web browser and allows users to mark problematic features from a set of predefined classes. 
Users can also generate custom labels and thus help identify previously unknown problem classes. 
User reports are fed back to hardware and software experts to help mitigate and eliminate recognized issues.
We report on the implementation of the application and our experience with its over 100 users, the majority of which are professional or prospective astronomers but not data management experts. 
We discuss aspects of user training and engagement, and demonstrate how problem reports have been pivotal to rapidly correct artifacts which would likely have been too subtle or infrequent to be recognized otherwise. 
We conclude with a number of important lessons learned, suggest possible improvements, and recommend this collective exploratory approach for future astronomical surveys or other extensive data sets with a sufficiently large user base. 
We also release open-source code of the web application and host an online demo version at \url{http://des-exp-checker.pmelchior.net}.
\end{abstract}


\end{frontmatter}

\section{Introduction}

Large astronomical surveys produce vast amounts of data for increasingly demanding science applications. 
At the same time, the complexity of the instruments, operations, and the subsequent data analyses renders glitches and flaws inevitable, particularly during the early phases of experiments. 
Thus, mechanisms that facilitate the discovery and reporting of problems in the data, whether they originate from unexpected instrumental behavior or insufficient treatment in software, are important components of a data quality program. 
Due to the often unexpected nature of these features, algorithmic approaches to identify artifacts are generally infeasible and human inspection remains necessary. 
Human-generated reports can then be fed back to algorithm developers and hardware experts to mitigate and eliminate problems whenever possible.
For current and upcoming surveys, the demands of carefully inspecting sizable portions of the data volume exceed the capabilities of individual, or a small team of, data management experts. 

Crowdsourcing has seen tremendous success in the past few years in many applications where a critical task cannot be performed by computers but where the amount of data to be gathered, processed, or analyzed exceeds the capabilities of even the most dedicated human. Examples can be found in non-profit, academic, commercial, or activist settings.
In astronomy, one of the first implementations of crowdsourcing was to gather information about the 1833 Leonid meteor storm \citep{Olmsted_1834a,Olmsted_1834b,Littmann_2014}. In recent years, widespread access to the internet has made such efforts easier to realize, allowing for larger crowds and quicker turnaround of results. The preeminent early adopter of this web-based mode of operations is the Galaxy Zoo project \citep{Lintott_2008}, designed to visually classify the morphology of galaxies from the Sloan Digital Sky Survey \citep{York_2000}. Galaxy Zoo has led to Zooniverse,\punctfootnote{\url{https://www.zooniverse.org/}} currently the largest online portal for citizen science projects. At the time of writing, Zooniverse findings have been published in over 80 articles across several disciplines of science.

We have built a web-based crowdsourcing application for image quality control for the Dark Energy Survey \citep[DES;][]{DES_2005}.\punctfootnote{\url{http://www.darkenergysurvey.org}} DES is a 5,000 deg$^2$ survey in five photometric bands ($grizY$) operating from the Blanco 4m telescope at the Cerro Tololo Inter-American Observatory (CTIO). 
Its 570 megapixel imager DECam \citep{Flaugher_2015} comprises 62 science CCDs (2048{\texttimes}4096 pixels) and 12 focus and guiding CCDs (2048{\texttimes}2048 pixels), covering a roughly hexagonal footprint of 3 deg$^2$. 
Each region of the survey footprint will be observed 10 times in each band over the course of five years. The total number of exposures is thus expected to be approximately $10^{5}$, with a data volume in science images of order 100 TB. An overview of the data processing and management pipeline (DESDM) is given by \citet{Mohr_2012} and \citet{Desai_2012}.

The application, dubbed the ``DES exposure checker'', is geared for a professional user base of several hundred scientists and seeks to identify flaws in the DES images that will be used for most science analyses. Problems discovered with this application can then be fixed in hardware or in subsequent data processing runs. Our approach ties in with other quality control efforts, which automatically analyze the latest exposures and flag cases of e.g.\ bad observing conditions \citep{Honscheid_2012, Diehl_2014}, and allow the inspection of final coadded data products, both images and object catalogs \citep{Balbinot_2012}.
Our concept profits from the experience of a prior ad hoc crowdsourcing effort in DES. During the so-called Science Verification phase in 2012, flaws in DES imaging have been identified by a small ``eyeball squad'', whose reports were relayed to DES operations and data management experts on a daily basis. While essential to improving the performance of the instrument during this early phase, the effort did not scale well with the increasing number of incoming images. In the remainder of this paper, we will show how to create a scalable solution to data quality control by providing an engaging user experience, while simultaneously maximizing the utility of the report collection.
The reports resulting from the DES exposure checker have already been used to inform the modification of existing algorithms, the development of new algorithms, and a general improvement the quality of the DES data.

\subsection*{Concept}
\label{sec:concept}

For many crowdsourcing applications, classification is the critical task performed by humans. Specifically, which class, of a predefined set, does a given test object belong to? To render this question accessible to as wide an audience as possible, the task is ideally broken up into yes-or-no questions. The approach we present here differs insofar as we ask our users not only to classify image flaws according to a given classification scheme, but also to extend the scheme if new classes of flaws arise. We can add this complexity because our audience consists of mostly experienced or prospective astronomers, who can contribute their prior knowledge to improve the classification scheme. This feedback system, catered to a professional---but not necessarily expert---user base, is the main novelty of our approach.
Specifically, we want to bring together three different objectives in one application:
\begin{enumerate}[label=(\arabic*), ref=(\arabic*), leftmargin=*, partopsep=0em, itemsep=0em ]
\item Show DES participants how science-grade images really look.\label{a:1}
\item Enable users to discover and classify image flaws of known or unknown type.\label{a:2}
\item Aggregate problem reports to improve data and processing quality.\label{a:3}
\end{enumerate}
\autoref{a:1} has an educational direction. To optimally exploit the data, participants should understand how the survey performs: what its capabilities and limitations really are. In addition, new participants, such as beginning students, should get a visual introduction to the current level of image quality. \autoref{a:1} also provides a substantial source of motivation. During the early phases of an experiment, surprising features may be present in the data. We therefore anticipate that participants are  eager to satisfy their curiosity, and we seek to streamline the application such that they can efficiently inspect the images.

Practical limitations are at the heart of \autoref{a:2}. The flaws we seek to identify stem from instrumental malfunctions, sub-par observational conditions, and, foremost, an insufficient treatment of various artifacts during the stages of data processing. Because that constitutes an extensive and poorly defined list of potential problems, human inspection remains irreplaceable. The key to \autoref{a:2} is to leverage the user's curiosity to perform this seemingly mundane task. Beyond curiosity, we expect---and our experience confirms---that users are genuinely motivated by a concern and a sense of responsibility for their respective scientific application. Consequently, different users care about different problems, and a diverse user base (according to scientific background, level of experience, etc.) is thus beneficial to identify different kinds of problems. The challenge of diversity lies in the difficulty in obtaining unequivocal classifications of known and especially previously unknown flaws.

\autoref{a:3} enhances the overall utility of the application. With the collection of problem reports, we seek to provide feedback to experts in charge of data management and developers of data reduction and analysis pipelines. We are able to provide a list of training cases to assess the performance of new or improved processing algorithms. In particular, with a sufficiently large user base we can generate rapid feedback if unexpected and even rare problems exists in any new data release.

The remainder of this article describes how we seek to combine these objectives. Although interconnected, we separate the topics broadly according to the three aspects listed above. In \autoref{sec:presentation} we describe the presentation of the image data and the main elements of the user interface; \autoref{sec:interaction} deals with user interaction and engagement; and \autoref{sec:results} summarizes our findings on how the application has been used and what improvements have been made to data processing in DES based on the problems our users identified. We conclude in \autoref{sec:conclusions}.

\section{Presentation}
\label{sec:presentation}

\begin{figure*}[ht!]
\includegraphics[width=\linewidth]{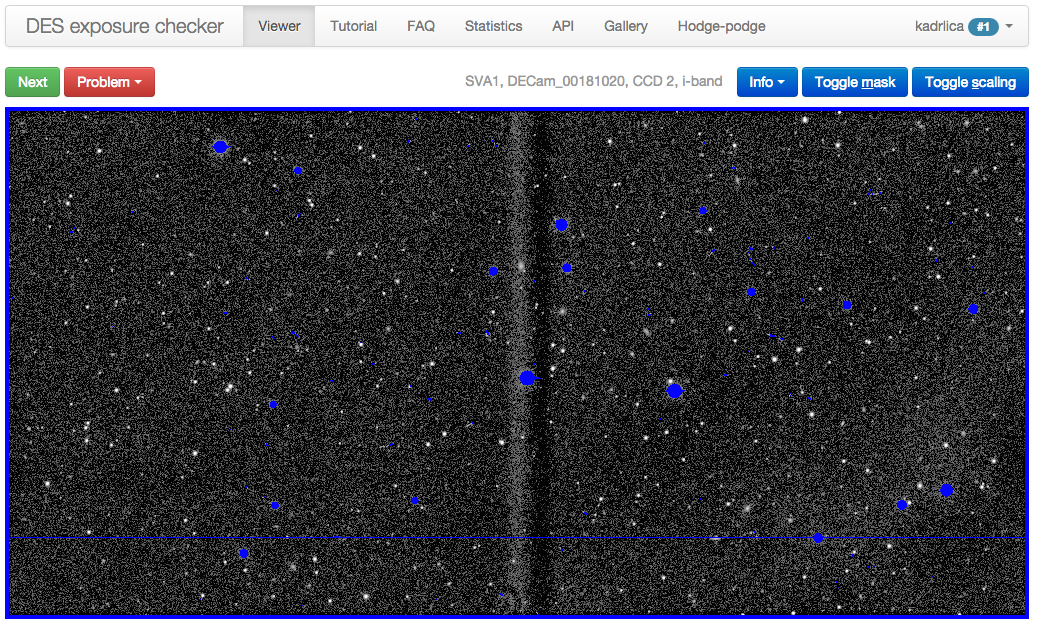}
\caption{\label{fig:viewer}
Screenshot of the DES exposure checker viewer page. It shows the fully reduced (i.e.\ science grade) FITS image of a single DECam CCD as delivered by an early version of the data processing pipeline. This image has been sky background subtracted and downsampled by a factor of four. Pixels colored in blue have been masked by the DESDM image processing pipeline. All elements of user interaction reside in the row above the image. The linear feature extending vertically accross the center of the image is the result of a shutter failure in the flat field for this exposure (\autoref{sec:vertical_jump}).%
}
\end{figure*}

Like many citizen science projects, we adopted a web-based approach to reach as many users (affiliated with the survey) as possible and to allow them to inspect images wherever and whenever they want. We also sought to eliminate installation hassles, which means that all features needed to be readily available without the need for extra software.

The web application is built with a SQLite database\footnote{\url{http://sqlite.org}} and PHP scripting\footnote{\url{http://php.net}} on the server side, with most of the functionality happening on the client side. The HTML page design derives from the Bootstrap framework\footnote{\url{http://getbootstrap.com}}, and essentially all interactive elements are written in JavaScript, utilizing the jQuery library.\punctfootnote{\url{http://jquery.com}}

To allow maximum flexibility in handling the images, we decided not to convert them into static images (e.g., in PNG format), but to load FITS files directly into the browser. For displaying the images in our viewer, we employ the pure JavaScript implementation WebFITS\punctfootnote{\url{https://github.com/kapadia/WebFITS}} \citep{Kapadia13.1}. This approach allows the users to alter properties of the visualization, such as image stretch, zoom, or the inclusion of overlay information, on the client side, without the need to reload additional data. Even when we deal with FITS files of only one of the 62 DECam CCDs, the full image comprises 4096{\texttimes}2048 pixels, which would exceed the size of virtually all commercial computer monitors and visually overwhelm the user. We therefore decided to compromise between providing the most authentic data unit, namely the entire unaltered image, and the most enjoyable user experience, which generally means reducing file sizes and load times. We decided to downsample the FITS image by a factor of 4 on the server side, reducing the image to 1024{\texttimes}512 pixels which, with compression, is about 450 kB in size. This choice yields images that conveniently fit into most browser windows and still provide enough information to spot even small flaws in the images. An example screenshot of the viewer page is shown in \autoref{fig:viewer}.

\begin{figure*}[h!]
\begin{center}
\includegraphics[width=0.864\linewidth]{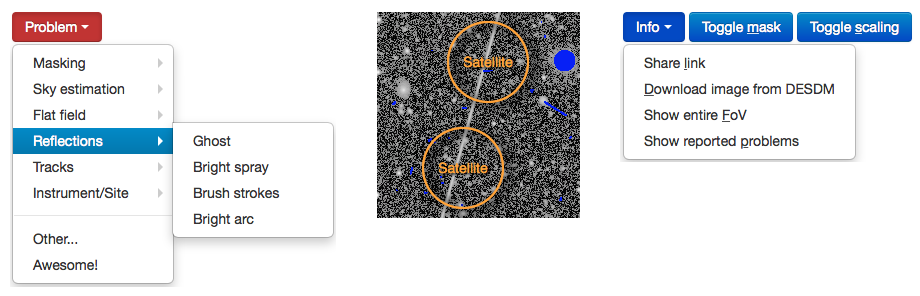}
\caption{\label{fig:viewer_details}
User interaction with the viewer page. \emph{Left:} Problem dropdown selector with open sub-menu. \emph{Center:} Cutout of a FITS image shown in the viewer with two problem markers (orange circles) indicating an earth-orbiting satellite trail (cf.\ \autoref{sec:satellites}). \emph{Right:} Info dropdown selector.%
}
\end{center}
\end{figure*}

The majority of the page is occupied by the image itself, with the elements of user interaction being confined to a single row above the image. At the most elementary level, only two buttons are needed. If users consider the image to be flawless, they click ``Next'' and are presented the next one. If they recognize flaws, they can select it from the dropdown menu labeled ``Problem'' (see left panel of  \autoref{fig:viewer_details}) and then click on the image to mark the location of the problem.

Additional information is provided on the right side, where we show the data release (``SVA1''), the unique exposure, CCD number, and the filter. Users can interact with the image by toggling on or off masked regions (shown as a blue overlay on the image). The masks, which are generated by DESDM to reject areas with e.g.\ bad columns or saturated stars, are stored in a second image extension in the FITS file and flattened to binary data format. In this process, any information as to why a pixel has been masked is lost, but the alternative, a multi-level mask selection or multi-color display of different masks, causes overly complicated visualizations while providing little actionable information. 

Finally, users can toggle the image stretch between settings that emphasize bright or faint features. Both use the same $\mathrm{arcsinh}$ scaling \citep[e.g.][]{Lupton_1999} but differ in the minimum intensity considered. The gray-scale value $v$ of a pixel with intensity $I$ is given by 
\begin{equation}
v = 255\ \frac{\mathrm{arcsinh}(I) - m}{M-m},
\end{equation}
where $M = \mathrm{arcsinh}(\max (I))$ and $m=\mathrm{arcsinh}(\min(I))$ for the faint or $m=\mathrm{arcsinh}(1)$ for the bright setting. Since the images are background-subtracted, i.e.\ the mean intensity of the sky is set to zero, the latter choice sets the majority of non-object pixels to black, while the former places visual emphasis on low-level noise.

With these two rather simple operations, users can quickly investigate different features of the image, e.g.\ whether an area is properly masked or structure is present in the noise. These interactions are rendered swiftly in the browser because the data are already fully loaded from the FITS file. We intentionally restricted user interactions with the image to maintain focus on the task at hand: namely to identify clearly visible problems. We therefore do not allow panning or zooming of the image, and only provide two image stretches and no color option.\punctfootnote{In a prototype, we explored a full-featured adjustment of the image stretch, which may improve the recognition of subtle problems at the expense of inspection time. A colored scheme for the image pixels instead of gray-scale would have increased the dynamic range but interferes with the colored problem markers, the visibility of which we considered critical.}

Once a problem has been selected from the dropdown menu, users can click on the image as often as they deem necessary to identify problematic features. If multiple problems are identified, they may be of the same or of different kind. We added two extensions to this scheme. To identify false positive problems, e.g.\ the application of a mask on an area that should not have been masked, we provide a ``False'' button to be pressed after the problem is selected. For problems of unknown type, the problem selector ``Other\ldots'' brings up a text box where users are asked to concisely describe what they see. The same workflow applies to the ``Awesome!'' selector for the case when users find a celestial body that is remarkable in its own right.

To support tablet computers, the act of marking a flaw is achieved with a single click or tap on the image. We decided to use circle markers so that we can place a problem label in their interior.\punctfootnote{We adhere to \citet{Jones_1989} and use the $\times$ symbol for false-positive reports, since $\times$ rarely, if ever, marks the right spot.} The marker and label color is yellow, which is both visually striking and clearly distinguishable from both bright and faint areas in the gray-scale image. A closeup of a marked area is shown in the center panel of \autoref{fig:viewer_details}. 

\begin{figure}[t!]
\begin{center}
\includegraphics[width=\columnwidth]{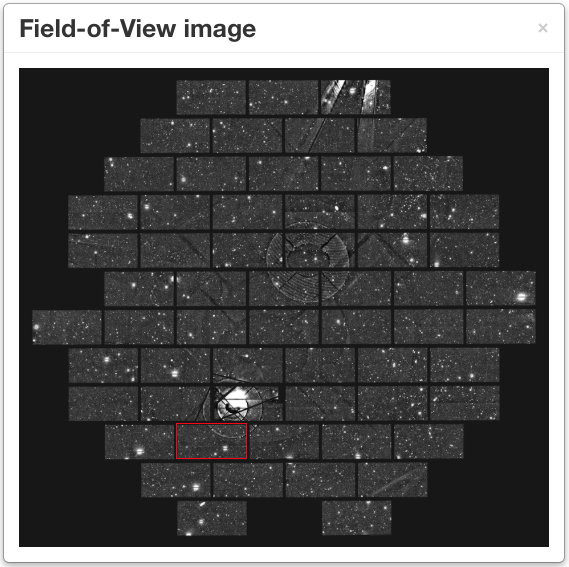}
\caption{\label{fig:fov}
Field-of-View visualization. Because many image artifacts (like the so-called ``ghosts'' caused by reflected light from bright stars, see \autoref{sec:scattered_light} for details) extend beyond a single CCD, it can be helpful to inspect the entire DECam array of a given exposure. The CCD image currently loaded in the viewer is highlighted by the red frame. Users can navigate to another CCD by clicking on its position in this image. Two CCDs are not shown; one failed early in the survey \citep{Diehl_2014}, and the other experiences a time-variable gain.
}
\end{center}
\end{figure}

Finally, users can request additional information about the image from the ``Info'' dropdown menu (see right panel of \autoref{fig:viewer_details}). We provide an unique URL for each image, so that users can share interesting cases, as well as the URL where the original FITS file (prior to our compression and background subtraction steps) can be downloaded for detailed inspection outside of the application. Because some problematic features span several CCDs and are easier to recognize when seen on a larger scale, we provide a static PNG image of the entire field of view (see \autoref{fig:fov}) with the option of switching to any other CCD of the same exposure. By selecting ``Show reported problems'', the viewer will render all problem markers it has received for the loaded image, thus allowing users to inform themselves what others users have found during their inspection. We exploit this dynamic problem lookup for user training in the \autoref{sec:consensus}.

\section{User interaction and engagement}
\label{sec:interaction}

The central design goal of the viewer page is to render the work convenient and to not frustrate the user:
\begin{itemize}[leftmargin=*, partopsep=0em, itemsep=0em ]
\item We sought to minimize the distance on screen and number of clicks to perform the most common tasks.
\item For changes in the visualization (such as image stretch, mask or problem overlay etc), we implemented keyboard shortcuts. 
\item Whenever the application loads information from the server, we show a spinner image to indicate a short wait time.\punctfootnote{While most queries to the server only require a few milli-seconds, retrieving a new FITS image can take up to one second. Without the spinner, the viewer would feel unresponsive during the load time, and without visual feedback users were found to click ``Next'' prematurely.}
\end{itemize}
Streamlining the viewer makes it less likely that users turn away from the application, but does not provide any motivation to use---or even continue to use---it in the first place.

\subsection{Motivation}
\label{sec:motivation}
\begin{figure}[t!]
\begin{center}
\includegraphics[width=\columnwidth]{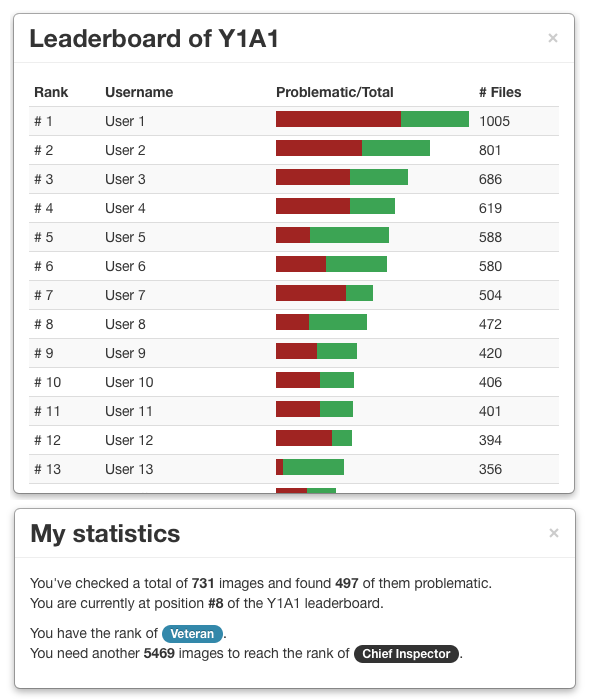}
\caption{\label{fig:gamification}
Elements of gamification. \emph{Top:} Leaderboard with rank, username (anonymized for the purpose of this publication), a stacked bar chart indicating the total number and the number of flawless images (green), and the image count for the given data release (``Y1A1''). \emph{Bottom:} User statistics listing the total number of images inspected (combining multiple data releases), the position on the release leaderboard, the current rank and, the number of images needed to attain the next-higher rank.%
}
\end{center}
\end{figure}

We pointed out in \autoref{sec:concept} that users involved in the survey have a genuine desire to understand the problems that may affect their science analyses and actively want to help improve the data quality. However, we sought to incorporate principles of gamification to provide additional motivation for users to keep using the application. Gamification results in a ``blurring of the relationship between work and play'' \citep[e.g.][]{Greenhill_2014}, and renders the task at hand into a competitive game that follows its own set of rules to define a winner. We employed a very simple metric to define success in our ``game'', namely the number of images users have inspected, irrespective of whether they found a problem or how many different problems they marked. The ranking of all users is shown as a leaderboard in the viewer (see top panel of \autoref{fig:gamification}), and the top 15 users are shown on the start page of the website as an incentive to rank above the cut. To prevent users from tricking the system by simply submitting empty problem reports---hitting ``Next'' repeatedly without inspecting the images---we resort to peer pressure. The leaderboard shows the fraction of images each user has considered flawless in green, triggering discussions among users when some of them appear ``too soft''.\punctfootnote{We do not provide any messaging system for users, but usernames are mostly recognizable and users know each other, so that email remained the preferred means of communication.}

We also implement a reward system with virtual badges. Based on the total number of images inspected, the user progresses through a rank system, with higher ranks becoming increasingly hard to attain:
\begin{itemize}[leftmargin=*, partopsep=0em, itemsep=0em ]
\item ``Rookie'' for 10 images
\item ``Frequent Checker'' for 62 images (corresponding to the number of DECam CCDs)
\item ``Top Performer'' for 5{\texttimes}62 images
\item ``Veteran'' for 10{\texttimes}62 images
\item ``Chief Inspector'' for 100{\texttimes}62 images
\end{itemize}
Whenever a user crosses one of the thresholds, a window will appear and congratulate the user for being awarded the next-higher rank. We chose a very low threshold for the lowest award to provide early gratification for most users. Once awarded, the colored badge is shown next to the username in the top-right corner of \autoref{fig:viewer}, highlighting the user's progress. Users also have access to their current statistics: the number of images they have inspected, their current rank, and the number of images they need to inspect for the next rank (see bottom panel of \autoref{fig:gamification}).

We furthermore created a dynamic gallery comprised of images flagged as ``Awesome'' in the viewer. Listing the username and timestamp together with the user's description of the object not only yields some bragging rights for the discoverer, it also allows other users to share the enjoyment of remarkable images of astronomical objects (a fascination that influenced many users to pursue careers in physics or astronomy). While most objects in the gallery are well-known nearby spiral galaxies, in a few cases users suspected that they had discovered a so-far unknown object. When this was the case, we attempted to verify the nature of the object in question. In one instance, we confirmed that a user had spotted a comet, \object{C/2010 R1}, that had been discovered only a few years earlier.

Finally, we engaged users by sending email newsletters that announced updates, such as the availability of a new data set or improvements to the application, and also by giving out bonus points for the leaderboard position whenever a user inspected 10 images during the days from December 1 to 24, 2013. Both approaches led to at least twice as many user contributions compared to a regular day.
  
\subsection{Consensus on problems}
\label{sec:consensus}

\begin{figure}[t!]
\begin{center}
\includegraphics[width=1\columnwidth]{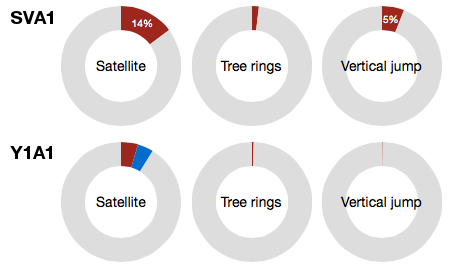}
\caption{\label{fig:statistics}
Aggregated statistics of problem reports, for the two first data releases of DES, SVA1 and Y1A1. Each donut chart shows the reported fraction of problematic images to exhibit this kind of flaw (red). False-positive fractions are colored in blue. See \autoref{sec:consensus} for a discussion of the reliability of the abundance estimates.%
}
\end{center}
\end{figure}

So far we have discussed the creation of an engaging application to interact with DES images, promoting \autoref{a:1}. To address \autoref{a:2} and \autoref{a:3} it is critical that users be able to find flaws, recognize and classify them, and even identify previously unclassified flaws. The primary element of user training is an extensively vetted tutorial page, which describes common problem classes and the best current understanding of their origins.\punctfootnote{We did not provide a facility to discuss the possible origin of artifacts as part of the tutorial, but rather relied on the DES collaboration wiki pages. These discussions were thus documented in a way that was familiar to survey participants and the conclusions were summarized in the tutorial.} Next to each description, we show a thumbnail of a characteristic example, which is linked to the viewer page, so that one can access the original FITS image and the problem markers reported by all users. Thus, the tutorial provides users with a visual reference of prototypical problem cases together with an understanding of their causes. The connection between cause and effect further satisfies the desire of survey participants ``to understand the data''.

In addition to the user tutorial, we implemented a statistic overview page, where problem reports were aggregated and listed for each problem class (a selection is shown in \autoref{fig:statistics}). We show how often each problem occurs, and the fraction of false positive problem reports for each class.\punctfootnote{In this context, false positive refers to cases where the processing pipeline masked an area of an image without any visible artifact.} The problem classes are linked to example images drawn at random from the images with corresponding problem reports. Beyond the prototypical cases shown in the tutorial, the random selection yields a dynamic impression of the variation between problem cases, or, more precisely, how they are perceived by users.

At this point we have to emphasize that these statistics do not necessarily provide a fair assessment of the overall data quality in DES. Many reports refer to insignificant flaws such as incompletely masked cosmic ray hits, which leave bright but only very short and thin streaks and therefore affect very small areas. Hardly any science application would be badly impacted by them, but we still count such an image as ``reportedly problematic''. Furthermore, the relative abundances of problem classes can become skewed towards the most obvious problems. For instance, large ghosts like those visible in \autoref{fig:fov}, often compel users to open the Field-of-View image and iterate through \emph{all} affected CCDs, marking the same ghost on several images, whereas they normally only inspect one CCD image of a given exposure if there is no large problem visible.
Also, with the overall data quality improving over time, users tend to shift their attention to subtler flaws, so that the aggregated statistics may not directly reflect the improvements to data quality that have actually been made.

Finally, we address the problem reports posted in the umbrella category ``Other'', chosen when users are uncertain about the type of problem. Similar to the gallery of ``Awesome'' objects, we list these cases together with the free-form description provided by the user on a page dubbed ``Hodge-podge''. Each problem report is linked to the corresponding FITS image in the viewer, and the page also shows how often the same description has been used on different images. The hodge-podge page is linked from the tutorial and constitutes a dynamic extension of the latter to currently unclassified problems. We ask users (as part of the tutorial and in email newsletters) to inspect the hodge-podge list to integrate these problems in their visual classification scheme and to adopt the existing description whenever they come across a fitting case.\punctfootnote{The description field in the viewer page pre-loads existing descriptions for swift selection and to avoid description mismatch e.g.\ from typos.} Because the list of hodge-podge problems is ranked according to their occurrence,  users are presented the most common---and hence, most important---problems first, similar to the concept of ranked comments on discussion websites such as \url{http://stackexchange.com}. 

This feedback loop, with which we dynamically integrate and extend the list of problem examples and even classes to take advantage of what users have reported, is the core element that distinguishes our approach from most crowdsourcing applications. For instance, ``Vertical jump'' was a common ``Other'' report early on and was therefore upgraded to a dropdown option (cf.\ \autoref{sec:vertical_jump}). We comment on the limitations of the current approach and how they can be remedied in \autoref{sec:conclusions}.
  
\section{Results}
\label{sec:results}

\subsection{Usage}

\begin{figure}[t!]
\begin{center}
\includegraphics[width=\columnwidth]{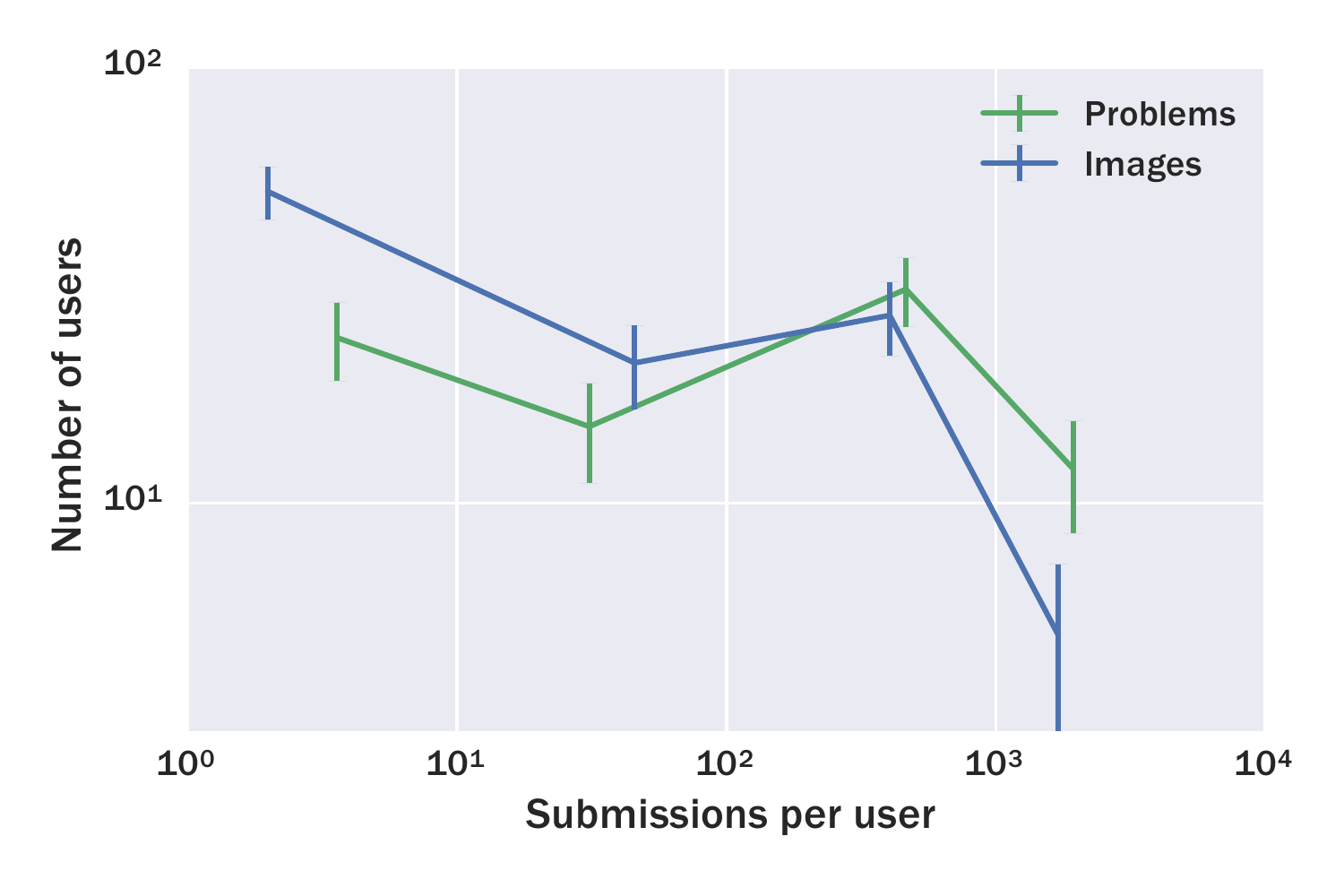}
\caption{\label{fig:user_stats}
Number of users within bins of either inspected images per user (blue) or total number of problems reported per user (green). The bins span one decade and are indicated by the vertical white lines, the horizontal position within each bin is given by its average submission number. Poissonian uncertainties are shown.%
}
\end{center}
\end{figure}

To date, 112 users have signed up on the DES exposure checker website, constituting a considerable fraction of the scientists and students in the DES Collaboration. For the first two data releases of DES, called SVA1 (with observations from 2012--2013) and Y1A1 (from 2013--2014), they have inspected a total of 21,147 images and submitted 39,213 reports.\punctfootnote{To facilitate inspection of the images and discussion within the collaboration, the viewer can be used without registration or the requirement to submit problem reports. The number of image views exceeds 250,000.}

As indicated in top panel of \autoref{fig:gamification}, only a few users are responsible for the majority of the submissions. \autoref{fig:user_stats} shows in more detail that about one half of the users inspected dozens or hundreds of images. When looking at the nearly flat distribution of number of problem reports per user (of which there can be several for each image), it becomes even more apparent that the most prolific users also attempted to be the most thorough.

To support studies of user reliability, we do not randomly present images to each user. Instead, we seek to maximize the number of images that have been inspected by at least three different users. Because a sizable fraction of the reports come from a small number of users, this cannot always be achieved, and the average number of distinct users per image is 1.58. Whenever multiple users inspected the same images, we found that they identified the same flaw(s) in 37\% of the images, with the most common form of disagreement being caused by more subtle flaws identified by one user but not by others. This type of incompleteness is not a major concern for achieving \autoref{a:3}, because any generated problem report will be conditioned on at least one user having found the problem in question. The second most common report mismatch stems from actual disagreement regarding a flaw's classification, followed by the rare case, in which two users identified two distinct and mutually exclusive problems. We refer to \autoref{sec:conclusions} for a discussion about the advantages of a user reliability assessment.

\subsection{Benefits for DES}
\label{sec:benefits}

Our main aim, besides informing survey participants, is to discover flaws in the data so that they can be corrected. Our effort is similar to the work of e.g.\ the ``LSC glitch group'', which visually monitored transients for LIGO and identified several instrumental causes for false alarms \citep{Blackburn_2008}. To facilitate the information exchange we provide an API to retrieve aggregated and anonymized reports for individual problems. Data processing experts can thus obtain lists of images and locations of problematic features to test and improve the quality of various image reduction and data processing steps. These lists are especially useful for issues that occur rarely, where it would take a single user a significant amount of time to assemble a sufficiently large data set.

\begin{figure*}[t!]
\begin{center}
\includegraphics[width=\linewidth]{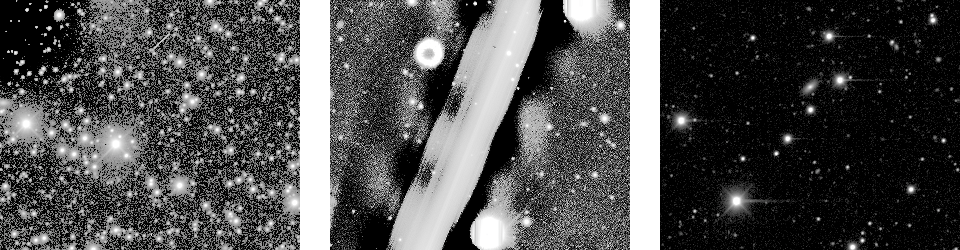}
\caption{\label{fig:problems}
Example problem cases. \emph{Left:} Sky background estimation was found to be problematic in dense stellar fields, particularly at the edges and corners of the CCDs (cf.\ \autoref{sec:background}). \emph{Center:} The light of airplanes ruins large portions of affected exposures (the image shown covers almost an entire CCD; cf.\ \autoref{sec:planes}). \emph{Right:} A small number of CCDs shows stars with one-sided trails, the cause of which is currently under investigation (cf.\ \autoref{sec:star_trails}).
}
\end{center}
\end{figure*}

\subsubsection{Shutter failure during flat field exposures}
\label{sec:vertical_jump}
 The first interesting finding based on problem reports is related to the exposure shown in \autoref{fig:viewer}, specifically the vertical banding feature in the center of the image. This was one of the first artifacts identified by users of the exposure checker website, and was found to be present in a large number exposures from the SV period of DES. The problem reports provided critical information for an investigation of the origin of the flaw. It was ultimately traced back to a number of flat field images that were read out while the camera shutter was partially open, causing a strong gradient in horizontal direction. Because several flat field images are combined to suppress the variance of the correction and the resulting so-called ``super-flat'' is then applied to all images taken over a period of approximately two weeks, the flaw was present in about 5\% of all images of the SVA1 data, making it relatively easy to find. However, the amount of information about the artifact and the speed with which that information became available through the exposure checker API, enabled a rapid re-design of the super-flat production as well as a change to the data acquisition that prevented the problem from occuring in subsequent data releases. As shown in \autoref{fig:statistics}, where this artifact is labeled ``Vertical jump'', the problem was prevalent in SVA1 but vanishes in Y1A1 after the aforementioned changes.

\subsubsection{Orbital satellite trails} 
\label{sec:satellites}
The DES exposures are 90 seconds in $g,r,i,z$-bands and 45 seconds in $Y$-band. During this time, earth-orbiting satellites can traverse a significant fraction of the DES focal plane leaving extended bright trails across multiple CCDs (a trail segment can be seen in the central panel of \autoref{fig:viewer_details}). Unlike the shutter failures, this artifact was expected before the implementation of the exposure checker. However, an algorithm had not been developed for identifying and masking this contamination. Humans excel at perceiving linear features in noisy data \citep[e.g.][]{Quackenbush_2004}, and the exposure checker users quickly identified a large sample of trails. While not specifically designed for this purpose, the  background subtraction and coarse binning we chose for the viewer proved beneficial for enhancing the contrast between satellite trails and the sky background.
The automated algorithm developed for satellite trail detection took advantage of the same background subtraction and binning steps employed by the exposure checker, and then applied a Hough transform \citep{Hough_1959,Duda_1972} for satellite trail detection and characterization. Due to the efficiency of the human eye in recognizing trails, the appropriate problem reports from the exposure checker was used as a truth sample for testing and optimizing the performance of the automated algorithm. Because we also ask our users to report false-positive satellite masks (cf.\ \autoref{fig:statistics}), the parameters of the algorithm were adjusted such that it does not over-aggressively trigger for proper elongated features like large elliptical galaxies. To date, the automated algorithm achieves a true-positive rate of $\sim$80\% with a false-positive rate of $\sim$2\%.\punctfootnote{Nearly all of the trails missed by the automated algorithm have a contrast of $<3\sigma$ with respect to the sky background. These trails are clearly visible to the human eye, but are algorithmically difficult to distinguish from elongated elliptical galaxies and large-scale imperfections in the background model.}

\subsubsection{SExtractor background estimation} 
\label{sec:background}
During investigation of the first data release, users noticed a systematic darkening of the sky background near the corners of many CCDs (see left panel of \autoref{fig:problems}). Initially labeled ``dark corners'' or ``dark boundary'', it was one of the most common problem reports from the hodge-podge list of that data release.  Occurring in crowded stellar fields (e.g.\ close to the Large Magellanic Cloud) or when bright stars happen to fall close to the boundaries of CCDs, further investigation revealed that these artifacts stem from the sky background estimation performed by SExtractor \citep{Bertin_1996}. Once identified, this issue was subsequently mitigated by the data processing team through adjustment of the SExtractor parameters.

\subsubsection{Airplanes} 
\label{sec:planes}
Two commercial Chilean flight paths pass close to CTIO. As opposed to bright orbital satellites which impact at most 10\% of the CCDs out of a full exposure, airplanes may render a large portion of the focal plane entirely unusable for scientific purposes (see central panel of \autoref{fig:problems}). While the occurrence of airplanes in the DES imaging is rare (several exposures per season), their tremendous impact makes it critical that affected exposures be identified and removed. Because airplanes can completely ruin a single-CCD image, the field-of-view visualization in the viewer page is particularly helpful to allow users to see their entire track across the DECam focal plane. Additionally, once an airplane is spotted in a single CCD, users can quickly navigate to other affected CCDs and tag them, too. Users identified four airplane tracks in the DES Science Verification data release and the affected CCDs were subsequently excluded from further analysis.
  
\subsubsection{Spurious reflections}
\label{sec:scattered_light}
While every care was taken to avoid scattering surfaces during the construction of DECam, some scattered light artifacts are visible when observing in the vicinity of bright stars. They appear as large-scale diffuse features with characteristic morphologies (e.g.\ the shape of the telescope pupil) originally identified by the eyeball squad and easily visible in exposure checker images. Since scattered light artifacts often span multiple CCDs, this is another problem class that benefits from the field-of-view visualization: often the characteristic patterns of scattered light artifacts are only clearly recognizable when multiple CCDs can be examined (cf.\ \autoref{fig:fov}).

Feedback from the exposure checker provided a test sample to validate automated masking algorithms, which can predict the location of the most intense reflections from a detailed ray-tracing simulation of the optical path, and to identify other sources of scattered light. Whereas some scattered light features are unavoidable,\punctfootnote{for example, the image ``ghosting'' from double reflections of light off of the CCDs despite their anti-reflective coating} others, such as reflections off of the walls of the filter changer, were mitigated by judicious application of anti-reflective paint once the cause was identified \citep[Section 6.3 in][]{Flaugher_2015}.

\subsubsection{Cosmic Rays}
\label{sec:cosmic_rays}
Thousands of cosmic rays interact with the DECam CCDs during each DES exposure.
Cosmic rays deposit excess charges into pixels leaving tracks that are much narrower than the point spread function of the image.
During SVA1 processing the cosmic-ray masking algorithm was known to not perform optimally.
This poor performance was independently confirmed by the results of the exposure checker. 
An improved masking algorithm was implemented for Y1A1 and the improved performance was validated by exposure checker users.
We caution that while users are clearly able to identify catastrophic failures in cosmic-ray masking, the small sizes and large numbers of cosmic-ray tracks make it unlikely that the reports are complete. This renders it difficult to quantify the detailed performance of an algorithm based on the exposure checker output alone.
  
\subsubsection{Star trails} 
\label{sec:star_trails}
An example of a relatively rare but visually recognizable feature is shown in the right panel of \autoref{fig:problems}. Dubbed ``horizontal star trails'', it has been identified by several users in 45 images spread over the SVA1 and Y1A1 releases. Our subsequent investigation revealed that all reports refer to images from only three CCDs, that these CCDs are far apart in the focal plane, and that neighboring CCDs of the same exposure are unaffected. The trails occur behind (in readout direction) moderately bright stars and are preferentially clustered on one half of the affected CCDs. A detailed investigation of this artifact is underway.

\section{Lessons learned}
\label{sec:conclusions}

Visual quality control is a feasible application of crowdsourcing. We operate under the assumption that participants in scientific experiments, such as the optical imaging survey DES, are genuinely interested in data flaws relevant to their science cases. Indeed, we found---and have been told---that if the interaction with the application is streamlined and frustration-free, our users enjoyed exploring the data, either to inspect the quality of new releases or, for new participants, to get to know their current state. Many users suggested improvements to the applications (e.g.\ the field-of-view visualization) that extended its capabilities and made the exploration more efficient and illuminating. We also found that users can be further motivated by newsletters, gaming incentives, such as badges and leaderboards, and follow-up investigations of unexpected discoveries. These findings all point to an engaged user base, and our usage statistics support such an interpretation.

By now, users have submitted more than 39,000 reports. Many of the reported flaws were initially not expected, so we created the umbrella class ``Other'', for which user-defined labels are required. By rank-ordering those labels, we could identify the most common types of flaws and extend our classification scheme to currently 26 distinct classes. The tutorial, which summarizes the classes and the current knowledge of their causes, has become an often-visited visual catalog of DES data flaws, and has clear educational value for survey participants. We furthermore provide an API to query anonymized problem reports. This database has been instrumental to identify the origin of several problem classes and to generate specific test cases for algorithms that mitigate or entirely eliminate the impact on subsequent science analyses.

Our findings are applicable to upcoming astronomical surveys and other extensive but error-prone data sets with a sufficiently large user base. We see clear benefits of the crowdsourcing approach to rapidly improve data quality in the early phases of an experiment, and to educate participants about its real-world limitations. In the light of growing data volumes and rates of forthcoming surveys such as LSST,\punctfootnote{\url{http://lsst.org}} we emphasize that a complete inspection of all images is not necessary to reveal various kinds of problems present in the data. When faced with truly overwhelming data volumes, we recommend pre-selecting images with potential problems for human inspection to keep the size of an effective user base within reasonable limits and to provide enough problems, so that the users remain motivated by their discoveries.

The application code currently operates with images in the astronomical FITS format but can be modified to work with any two-dimensional data representation. A demo version of the application as employed by DES is hosted online\footnote{\url{http://des-exp-checker.pmelchior.net}}, and we publicly release the source code.\punctfootnote{\url{https://github.com/pmelchior/des-exp-checker}}

\subsection*{Potential improvements}

\begin{itemize}
\item While we recommend that users read the tutorial, there was no formal training process. To establish a more informed and confident user base, an interactive introduction could be made mandatory, wherein each new user is presented with several thoroughly vetted problem cases of increasing difficulty. While such a requirement could be perceived as onerous labor and might reduce the number of active users from the beginning, it would likely improve the classification accuracy. Following \citet{Marshall16.1}, we thus recommend a short and engaging training session for new users to even out differences in professional experience.
\item The number of recognized problem classes has by now grown to 26. To help users during classification, prototypical images from the tutorial could be made available on the viewer page as a swift visual index of the classification, for instance in the form of a collapsible side bar.
\item To maximize the utility of the problem reports with respect to \autoref{a:3}, i.e.\ to help inform the improvement of data processing algorithms, we need problem classifications with small type-I error, whereas type-II errors are irrelevant.\punctfootnote{Type-I and type-II errors are often referred to as ``false positives'' and ``false negatives'', respectively. In astronomical terms, we seek high purity of the classification at any level of completeness.} In this context, it is important to note that most mismatches in problem reports stem from type-II errors, whereas cases where users identify the same flaw but classify it differently (thereby causing at least one type-I error) are less frequent. Nonetheless, type-I errors could be reduced if we knew how often they occur for any given user, which would allow us to weigh reports according to the probability $p(c\,|\,u)$ of any user $u$ correctly classifying the problem of interest $c$. \citet{Lintott_2008} suggested a scheme, in which the agreement with the majority classification for objects with multiple submissions from different users determines the user weights. Because it is difficult to produce sufficient redundancy with a smaller user base, we prefer an alternative proposed by \citet{Marshall16.1}: The necessary information can be gathered by occasionally inserting training images into the normal stream presented in the viewer and tracking how often users correctly identify flaws. We note that verifying whether a user has found a pre-determined problem is a non-trivial task for problem classes with large spatial extent (like ghosts or reflections): In contrast to simpler distance-based success criteria, complex polygonal shapes would be required for such training images to determine if the markers were placed within respective target areas.
\item The random insertion of training images would also help to re-calibrate the overall problem statistics, i.e.\ the total number $N(c)$ of problem reports for class $c$ as shown in \autoref{fig:statistics}. One can imagine a case where a given user is rather impervious to a problem class, biasing its reported abundance low, or a case of a thorough user who inspects neighboring CCDs when they might be affected by the same problem, causing the abundance to be biased high. By determining how often users report a particular problem when it has been presented to them in a hidden test, $N_\mathrm{t}(c\,|\,u)$, one can estimate a corrected abundance as $\sum_u N(c\,|\,u) / N_\mathrm{t}(c\,|\,u)$.\punctfootnote{Note that for the former case $N_\mathrm{t}(c\,|\,u) = p(c\,|\,u)$, whereas the latter case requires that reports of the same class for images inspected by the same user directly after the test images need to be combined.}
\item The hodge-podge list of unknown problem classes currently ranks new cases by their total occurrence. Hence, type-I errors pose the same kind of problem as for the pre-determined classification scheme, namely if one user mistakenly starts to identify some aspect of the image as a new class of problems when it is a known class or no problem at all. The ranking could be improved if the problem reports are weighed with $\sum_c p(c\,|\,u)$, i.e.\ the probability of the submitting user $u$ correctly classifying \emph{all} existing problems.
\item An alternative mechanism to rank the hodge-podge list, potentially in addition to simple or weighted occurrence, is user voting. By giving each user a vote to either agree or disagree with the existence of the proposed problem class, one could reduce the impact of poorly informed or even rogue users who continue to assign a particular description to a problem that is either already known or non-existent.\punctfootnote{Due to the small user base of professional collaborators who mostly know each other, we never experienced intentionally malicious behavior.} 
\item It could be beneficial to discuss if an image is problematic or what kind of flaw can be observed directly within the viewer, e.g.\ in a discussion thread below the image, instead of on a separate wiki page. For instance, in case of uncertain classifications from the hodge-podge list, experienced users could provide their comments and help clarify whether the flaw is of a new kind. The downside of such a solution is that these discussions would be spread out over many distinct images, whereas a wiki page provides a central location for a coordinated discussion. Such a downside could be overcome through the implementation of a discussion forum where individual images could be discussed and two-way references could be formed between images and forum topics.
\item The gallery of ``Awesome'' objects comprises many of the most extended or otherwise remarkable objects of the survey. An automated script that queries our API and creates multi-color images of all objects on the awesome list would visibly recognize the effort of the users and provide a valuable resource for outreach purposes.
\end{itemize}
We encourage any interested party to contact us with further suggestions of improvements through the issue tracker system of the code repository.

\section*{Acknowledgments}
\label{sec:acknowledgments}

\hyphenpenalty=0
\exhyphenpenalty=0

We thank high school student Tawny Sit, who, while working at BNL, helped to identify and categorize problem cases that did not yet have a previously dedicated category.

We are grateful for the extraordinary contributions of our CTIO colleagues and the DECam Construction, Commissioning and Science Verification teams in achieving the excellent instrument and telescope conditions that have made this work possible.  
The success of this project also relies critically on the expertise and dedication of the DES Data Management group.

Funding for the DES Projects has been provided by the U.S. Department of Energy, the U.S. National Science Foundation, the Ministry of Science and Education of Spain, 
the Science and Technology Facilities Council of the United Kingdom, the Higher Education Funding Council for England, the National Center for Supercomputing 
Applications at the University of Illinois at Urbana\-/Champaign, the Kavli Institute of Cosmological Physics at the University of Chicago, 
the Center for Cosmology and Astro\-/Particle Physics at the Ohio State University,
the Mitchell Institute for Fundamental Physics and Astronomy at Texas A\&M University, Financiadora de Estudos e Projetos, 
Funda{\c c}{\~a}o Carlos Chagas Filho de Amparo {\`a} Pesquisa do Estado do Rio de Janeiro, Conselho Nacional de Desenvolvimento Cient{\'i}fico e Tecnol{\'o}gico and 
the Minist{\'e}rio da Ci{\^e}ncia, Tecnologia e Inova{\c c}{\~a}o, the Deutsche Forschungsgemeinschaft and the Collaborating Institutions in the Dark Energy Survey.

The Collaborating Institutions are Argonne National Laboratory, the University of California at Santa Cruz, the University of Cambridge, Centro de Investigaciones Energ{\'e}ticas, 
Medioambientales y Tecnol{\'o}gicas Madrid, the University of Chicago,
University College London, the DES\-/Brazil Consortium, the University of Edinburgh, 
the Eidgen{\"o}ssische Technische Hochschule (ETH) Z{\"u}rich,
Fermi National Accelerator Laboratory, the University of Illinois at Urbana\-/Champaign, the Institut de Ci{\`e}ncies de l'Espai (IEEC/CSIC), 
the Institut de F{\'i}sica d'Altes Energies, Lawrence Berkeley National Laboratory, the Ludwig\-/Maximilians Universit{\"a}t M{\"u}nchen and the associated Excellence Cluster Universe, 
the University of Michigan, the National Optical Astronomy Observatory, the University of Nottingham, The Ohio State University, the University of Pennsylvania, the University of Portsmouth, 
SLAC National Accelerator Laboratory, Stanford University, the University of Sussex, and Texas A\&M University.

The DES data management system is supported by the National Science Foundation under Grant Number AST\-/1138766.
The DES participants from Spanish institutions are partially supported by MINECO under grants AYA2012\-/39559, ESP\-2013\-/48274, FPA2013\-/47986, and Centro de Excelencia Severo Ochoa SEV\-/2012\-/0234.
Research leading to these results has received funding from the European Research Council under the European Union{'}s Seventh Framework Programme (FP7\-/2007\-/2013) including ERC grant agreements 240672, 291329, and 306478.

PM is supported by the U.S. Department of Energy under Contract No. DE-FG02\-/91ER40690.
ES is supported by DOE grant DE-AC02\-/98CH10886.
This work was supported in part by the National Science Foundation under Grant No. PHYS\-/1066293 and the hospitality of the Aspen Center for Physics.

This paper has gone through internal review by the DES collaboration.

\section*{References}

\bibliography{bibliography.bib}

\end{document}